\documentstyle[preprint,12pt,aps,psfig]{revtex}

\tighten
\begin{document}
\draft
\preprint{Imperial/TP/96-97/21}

\newcommand{\nc}{\newcommand}
\nc{\al}{\alpha}
\nc{\ga}{\gamma}
\nc{\de}{\delta}
\nc{\ep}{\epsilon}
\nc{\ze}{\zeta}
\nc{\et}{\eta}
\renewcommand{\th}{\theta}
\nc{\ka}{\kappa}
\nc{\la}{\lambda}
\nc{\rh}{\rho}
\nc{\si}{\sigma}
\nc{\ta}{\tau}
\nc{\up}{\upsilon}
\nc{\ph}{\phi}
\nc{\ch}{\chi}
\nc{\ps}{\psi}
\nc{\om}{\omega}
\nc{\Ga}{\Gamma}
\nc{\De}{\Delta}
\nc{\La}{\Lambda}
\nc{\Si}{\Sigma}
\nc{\Up}{\Upsilon}
\nc{\Ph}{\Phi}
\nc{\Ps}{\Psi}
\nc{\Om}{\Omega}
\nc{\ptl}{\partial}
\nc{\del}{\nabla}
\nc{\be}{\begin{eqnarray}}
\nc{\ee}{\end{eqnarray}}
\nc{\lambar}{\overline{\lambda}}

\title{Non-perturbative Expansion, Renormalons, and $\tau$ Decay}
\author{H. F. Jones$^a$\footnote{email: h.f.jones@ic.ac.uk}, 
        A. Ritz$^a$\footnote{email: a.ritz@ic.ac.uk}, 
        and I.L. Solovtsov$^b$\footnote{email: solovtso@thsun1.jinr.dubna.su}
        \\ $\;$\\}
\address{$^a$ Theoretical Physics Group, Blackett laboratory, \\
         Imperial College, Prince Consort Rd., 
         London, SW7 2BZ, United Kingdom.\\$\;$\\}
\address{$^b$ Bogoliubov Laboratory of Theoretical Physics,
         Joint Institute for Nuclear Research, \\
         Dubna, Moscow Region, 141980, Russia \\ $\;$\\}

\maketitle

\begin{abstract} 
 We analyze inclusive $\tau$ decay using a modified version of the $a$
expansion, a non--perturbative technique in which the effective
coupling is analytic in the infrared region. 
The modification involves renormalization group improvement of the 
integrand in a spectral representation for the $D$--function prior to 
implementing the $a$ expansion.  
The advantage of this approach is that it enables 
us to monitor the structure of the induced
power corrections and to ensure that these are consistent with the
operator product expansion.  
Numerically the method agrees well with
experiment: the comparison is made with the physical quantity $R_Z$
using $R_\tau$ as input.
\end{abstract}
        
\pacs{PACS Numbers : 13.35Dx, 11.10Hi, 12.38Lg} 

\section{Introduction}
Inclusive $\tau$--decay provides an ideal arena
for extracting information about the low energy domain of the strong
interactions, in particular the QCD coupling constant $\al_s(M_{\ta}^2)$.
A detailed theoretical analysis of this process 
has been presented in \cite{braaten92} (see also 
\cite{braaten88,braaten89,lediberder92,truong93,chetyrkin93,altarelli95b,narison95,pich95b}).
There has been recent interest in trying to estimate
the intrinsic uncertainty in such results due to the finite order truncation
of the perturbative series. Attempts at going beyond this 
limitation have included the use of analytic continuation
to resum the so-called $\pi^2$--corections 
\cite{pennington81,krasnikov82,bjorken89,gorishny91,pivovarov91,pivovarov92a,pivovarov92b,groote97}, and infrared renormalon
resummations \cite{altarelli95,neubert95b,ball95} (see also 
\cite{zakharov92,beneke92,beneke93,beneke94,bigi94,beneke95,neubert95,maxwell95,maxwell96}). 

The starting point for the theoretical analysis is the expression
\cite{braaten92}
\be
 R_{\ta} &=& \frac{\Ga(\ta^- \rightarrow \nu_{\ta}{\rm hadrons}(\ga))}
               {\Ga(\ta^- \rightarrow \nu_{\ta}e^-\overline{\nu}_e(\ga))}
      = 2\int_0^{M_{\ta}^2}\frac{{\rm d}s}{M_{\ta}^2}
   \left(1-\frac{s}{M_{\ta}^2}\right)^2\left(1+\frac{2s}{M_{\ta}^2}\right)
     \tilde{R}(s), \label{rtau}
\ee
where $\tilde{R}(s) = Im \Pi_{A,V}(s+i\ep)/\pi$
\footnote{All definitions are as in \cite{jones95c}, and we use the 
standard convention $Q^2=-q^2$ where $q$ is the current momentum.} 
(see e.g. \cite{braaten92}).
The integral is not amenable to standard perturbation theory as it
runs over small values of momentum. However, using Cauchy's theorem, 
one may rewrite
this result as a contour integral in the $s$--plane with a contour
running clockwise round the circle $|s|=M_{\ta}^2$,
\be
 R_{\ta} & = & \frac{1}{2\pi i}\oint_{|z|=1} \frac{{\rm d}z}{z}
        (1-z)^3(1+z)\tilde{D}(M_{\ta}^2z), \label{cont}
\ee 
which naively
appears to avoid the low momentum region. However, this trick requires
the Adler $D$--function, $D(Q^2)=-Q^2d\Pi(Q^2)/dQ^2$, to have specific
analytic properties, namely to be an analytic function in the 
$Q^2$--plane except for a cut along the negative real axis. 
This property is broken in 
perturbation theory due to the presence of the Landau pole
at $Q^2=\La_{QCD}^2$. Consequently all perturbative treatments
are sensitive to the prescription for dealing with the Landau pole.

For this reason, in a previous publication \cite{jones95c},
an analysis of $\ta$--decay was presented making use of an
effective running coupling constant which is infrared (IR) finite and respects
the above mentioned analytic properties \cite{solov94a,solov94b}. While
numerically quite successful, a conceptual problem with this 
approach was that one had little knowledge of the structure of induced
non-perturbative contributions, in particular power corrections.
In this letter we present a revised formalism which, by first
analysing the resummmation ambiguity of the perturbative
series, allows us much greater control over the induced power
corrections. In Section 2 we explain this modification, and in
Section 3 apply it to the analysis of $\ta$--decay, extracting
a value of $R_Z$ by running the effective coupling up to the $Z$--scale.
Conclusions and future applications of this technique are discussed
in Section 4.

\section{The $a$--Expansion Formalism}
The nonperturbative expansion method proposed in \cite{solov94a,solov94b}
(see also \cite{jones95a,solov95})
allows one to systematically study the low energy regime of QCD
and evaluate the integral (\ref{rtau}) either directly or via use
of Cauchy's theorem as in (\ref{cont}) \cite{jones95c,jones95b}. 
The method is based on a new expansion
parameter $a$ connected with $\al_s$ by the equation
\be
 \la &\equiv& \frac{\al_s}{4\pi} \equiv \frac{g^2}{(4\pi)^2}
      = \frac{1}{C}\frac{a^2}{(1-a)^3}, \label{lam}
\ee
where $C$ is a positive variational parameter fixed via data from meson
spectroscopy \cite{solov94b}. An arbitrary Green function
may be expanded as a power series in $a$, a result achieved in
practice by a resummation of a truncated perturbative
series, replacing $\al_s$ by $a$ via the appropriate expansion
of (\ref{lam}). It is clear from (\ref{lam}) that for all positive values
of $\la$ the parameter $a$ lies in the region $0\leq a <1$.

The $Q^2$ evolution of $a$ is defined by the renormalization
group (RG) equation,
\be
 \label{nprge}
 f(a) &=& f(a_0) + \frac{2\beta_0}{C} \ln \frac{Q^2}{Q^2_0},
\ee
where $a_0=a(Q_0^2)$. We shall work at $O(a^5)$ which, from the
expansion of (\ref{lam}), allows the use of perturbative results at
$O(\la^2)$. Calculation of the non-perturbative $\beta$--function
at this order \cite{solov94b} leads to
\be
 f(a) & = & \frac{1}{5(5+3B)}\sum_{i=1}^3 x_i J(a,b_i),
\ee
where
\be
 J(a,b) & = & -\frac{2}{a^2b}-\frac{4}{ab^2}-\frac{12}{ab}
       -\frac{9}{(1-a)(1-b)}+\frac{4+12b+21b^2}{b^3}\ln a \nonumber\\
  &  & \;\;\;\;\;\; +\frac{30-21b}{(1-b)^2}\ln(1-a)
      -\frac{(2+b)^2}{b^3(1-b)^2}\ln(a-b).
\ee
In this expression 
\be
 x_i & = & \frac{1}{(b_i-b_j)(b_i-b_k)},
\ee
where $ijk=123$ and cyclic permutations, the values of 
$b_i$ are the solutions of the equation 
$1+ 9a/2 + 2(6+a)a^2 + 5(5+3B)a^3=0$, where $B=\beta_1/(2C\beta_0)$, and
$\beta_0$ and $\beta_1$ are the perturbative 1-- and 2--loop
coefficients of the $\beta$--function.
 
For subsequent analysis of $\ta$--decay it is useful to separate the higher
order terms of the $D$--function via the definition
\be
 D(Q^2) & = & 3\sum_f Q_f^2 (1+4\la_{eff}(Q^2)), \label{dfn}
\ee
where one may consider the contribution to $\la_{eff}$ either from
perturbation theory or from the $a$--expansion as appropriate. A direct
application of the latter to $\ta$--decay, as in \cite{jones95c,jones95b}, 
then corresponds to
defining $\la_{eff}$ via the appropriate expansion of (\ref{lam}).
However, the conceptual difficulty with this procedure is 
that one has little knowledge
of the induced non-perturbative corrections contained in the
variational series. In order to have control over this aspect we must
first analyse the structure of the RG improved perturbative series.

To do this in a manner which respects the analytic
properties of the $D$--function we use the standard spectral representation,
\be
 \label{disp}
 D(Q^2,\la) & = & Q^2 \int^{\infty}_{0} {\rm d}s \frac{1}{(s+Q^2)^2} R(s,\la),
\ee
At next--to--leading--log (NLL) order $R(s)$ is given by \cite{gorishny86}
\be
 R(s) & = & 3\sum_f Q^2_f \left[1+4\la+\left(a_1-a_2\ln\frac{s}{\mu^2}
            \right)\la^2\right],
\ee 
where
\be
 a_1 &=& \frac{2}{3}\left[3365-22f-8\zeta(3)(33-2f)\right], 
         \;\;\;\;\;\;\;\;\;\;
        a_2\;=\;4\beta_0,
\ee
It is now convenient to perform an integration by parts in (\ref{disp}),
which results in the expression
\be
 \la_{eff}(t,\la) & = & \int^{\infty}_0 {\rm d}\ta \om(\ta) 
      \left[\la+\frac{1}{4}(a_1+a_2-a_2\ln(t\ta))\la^2\right], \label{leff2}
\ee
where $t=Q^2/\mu^2$ and we have used (\ref{dfn}). The weight function 
$\om(\ta)$, given by
\be
 \om(\ta) & = & \frac{2\ta}{(1+\ta)^3}, \label{virtdis}
\ee
describes the distribution of virtuality. One may now use
RG improvement under the integral, first discussed in \cite{ginzburg66},
with the knowledge that $R(s)$ obeys the same homogeneous RG
equation as $D$. Using the two--loop $\beta$--function
$\beta(\la)=-\beta_0 \la^2-\beta_1 \la^3$, at NLL order we obtain
\footnote{In Eq. (\ref{lameff}), we have restored the factor $k$ representing
the renormalization scheme dependence. In the so--called V--scheme
$k_V$=1, and in the $\overline{MS}$--scheme which we shall use throughout
$k_{\overline{MS}}=\exp(-5/3)$ (see, e.g. \cite{neubert95b}).}
\be
 \label{lameff}
 \la_{eff}(t,\la) & = & \int^{\infty}_0 {\rm d}\ta \om(\ta) 
      \left[\frac{\lambar}{1+\beta_0\lambar\ln\ta}
       - \frac{\beta_1}{\beta_0}\lambar^2
       \frac{\ln(1+\beta_0\lambar\ln\ta)}{(1+\beta_0\lambar\ln\ta)^2}
       + d\frac{\lambar^2}{(1+\beta_0\lambar\ln\ta)^2}\right],
\ee
where $d=(1/4)(a_1+a_2(1-\ln k))$ and $\lambar\equiv\la(t)$. This
expression is now in a form which on inspection will exhibit the divergences
associated with large orders in perturbation theory. We observe that the
virtuality distribution function (\ref{virtdis}) exactly coincides with the 
function used in \cite{bigi94} and is numerically very close to that
obtained in \cite{neubert95} (see Figure 1), which in contrast to the
present construction was obtained via an all--orders 
resummation of renormalon contributions in the large $\beta_0$ limit.

\begin{figure}
 \centerline{%
   \psfig{file=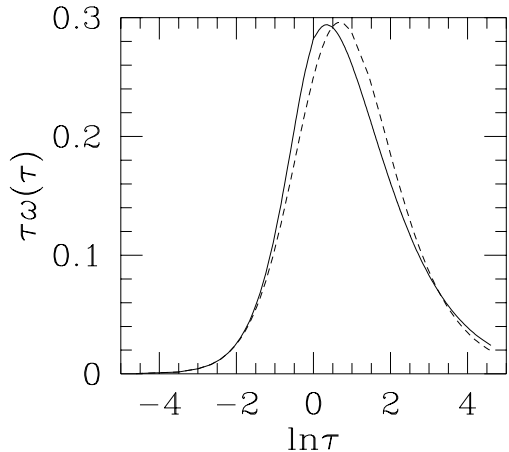,width=7cm,angle=0}%
   }
 \vspace*{0.2in}
 \caption{The virtuality distribution functions
  $\ta\,\om (\ta)$ taken from Ref.~\protect\cite{neubert95} 
  (solid line) and the
  function (\protect\ref{virtdis}) multiplied by a factor of 
  $\ta$ (dashed line) versus
  $\ln \tau$.}
        \label{virt}
\end{figure}

This connection with renormalons is illustrated more clearly by 
performing a formal Borel transformation on (\ref{lameff}), whereby
we obtain
\be
 \la_{eff}(t,\la) & = & \int_0^{\infty} {\rm d}b 
      \exp\left(-\frac{b}{\lambar(t,\la)}\right)B(b),
\ee
with
\be
 B(b) & = & \left[1+\left(\frac{\beta_1}{2\beta_0}+2d\right)b\right]
                \Ga(1+b\beta_0)\Ga(2-b\beta_0) + \frac{\beta_1}{4\beta_0}b.
\ee
This Borel function exhibits the correct infrared and 
ultraviolet renormalon poles for the $D$--function, although not the full
branch structure \cite{mueller85,broadhurst93}. Nevertheless, the fact
that the poles are correctly positioned implies that that the 
resummation ambiguity associated with the first IR renormalon
is $O(1/Q^4)$, which is consistent with the lowest dimension
vacuum condensate operator in the operator product expansion (OPE)
for the $D$--function. \footnote{Note that the partial integration 
performed to obtain
(\ref{leff2}) which generates the appropriate virtuality 
distribution is also responsible 
for the removal of the first IR singularity at $b=1/\beta_0$
required for consistency with the OPE.}

This result is important, as we may now conclude that introduction
of the $a$--expansion at this stage, given convergence of the
variational series, as has been proven in simpler systems
\cite{buckley93,guida95,arvan95c}\footnote{By convergence we mean strictly 
absolute convergence of the sequence of approximants for the $D$--function
$\{D_1(O(a),C_1),D_2(O(a^2),C_2),\ldots,D_N(O(a^N),C_N)\}$.}, may 
only induce those non-perturbative power corrections
required to cancel this perturbative ambiguity\footnote{Note that a 
$1/Q^2$ correction may also be induced in the
running coupling by removal of the Landau pole \cite{ball95}. However
this has a purely perturbative origin.}. Thus,
from the discussion above these corrections will only start at
$O(1/Q^4)$ and will therefore be consistent with an OPE treatment.

Introducing the $a$--expansion for the $D$--function in this 
modified spectral representation leads to the replacement of 
(\ref{lameff}) by
\be
 \label{lteff}
 \la_{eff} & = & \int_{0}^{\infty}{\rm d}\ta \om(\ta) \tilde{\la},
\ee
where to $O(a^5)$,
\be
 \tilde{\la} & = & \frac{a^2}{C} + \frac{3a^3}{C} + \frac{a^4}{C}
     \left[6+\frac{d}{C}\right]
      +\frac{a^5}{C}\left[10+6\frac{d}{C}\right].
   \label{ltilde}
\ee
The running expansion parameter $a=a(Q^2)$ is determined 
via (\ref{nprge}), and the variational parameter has been
determined as $C=21.5$ at $O(a^5)$ (and also as $C=4.1$ at $O(a^3)$)
\cite{solov94b}. The $D$--function constructed in this way may now
be utilised in an analysis of $\ta$--decay.

\section{$\ta$--Decay Analysis}
Making use of the contour integral representation (\ref{cont}) we isolate
the QCD contribution by defining 
$R_{\ta} = R_{\ta}^{(0)}(1+\De R_{\ta})$, with
\be
 R_{\ta}^{(0)} & = & 3(|V_{ud}|^2+|V_{us}|^2)S_{EW}.
\ee
The electroweak factor and the CKM matrix elements are
$S_{EW}=1.0194$, $|V_{ud}|=0.9753$, and $|V_{us}|=0.221$ respectively,
taken from \cite{braaten92}. Using the relations (\ref{dfn})
and (\ref{lteff}), and Cauchy's theorem, 
we obtain\footnote{In this case we take 
the u,d, and s quarks to be massless and thus 
ignore threshold effects.}
\be
 \De R_{\ta} & = & 48\int_0^{M_{\ta}^2} \frac{{\rm d}s}{M_{\ta}^2}
        \left(\frac{s}{M_{\ta}^2}\right)^2 \left(1-\frac{s}{M_{\ta}^2}\right)
        \tilde{\la}(ks), \label{delr2}
\ee
which has a modified kinematic factor as compared to the standard
relation (\ref{cont}) with the maximum shifted to $s = (2/3)M_{\ta}^2$.

Extracting the experimental value of $R_{\ta}$ from \cite{pdg96},
as described in \cite{pich95b}, we obtain $R_{\ta}^{exp} = 3.64 \pm 0.02$.
Using this as input we obtain the effective coupling 
$\al_s(M_{\ta}^2)=0.31 \pm 0.01$ at $O(a^5)$. This result is not directly
comparable with perturbative extractions of $\al_s$, as it also
includes various non--perturbative corrections, and a correction
due to the removal of the Landau pole. Nevertheless we note
that this quantity is lower than most extractions in fixed order
perturbation theory, and that this shift is consistent with
expectations from renormalon resummmations \cite{ball95,girone96}.

It is more consistent to consider only physical quantities, and 
therefore we use the experimental result 
for $R_{\ta}$ to run the coupling up to the
$Z$--scale and extract $R_Z$. In order to evaluate this quantity we
apply the matching procedure in the physical timelike channel where,
at least to leading order, the change in the number of active quarks is
easily associated with the threshhold for pair production. The effective
coupling and its derivative are required to be continuous at the threshold
points, which is acheived by solving the continuity equations for
$C_f$ and $a_0^f$. Using the standard heavy quark masses,
$m_c=1.6$ GeV and $m_b=4.5$ GeV, and applying this matching
procedure we find
\be
  R_Z & = & 20.96 \pm 0.01\;\;\;\;\;\;\;\;\mbox{at }\;O(a^3),\\
      & = & 20.83 \pm 0.01\;\;\;\;\;\;\;\;\mbox{at }\;O(a^5),
\ee
which, at $O(a^5)$, is within one standard deviation of the experimental result
$R_Z=20.77 \pm 0.07$ \cite{pdg96}. Note that this is a considerably
better fit to the data than using the technique of \cite{jones95c}
when one accounts for the change in the data over the intervening
period.

\section{Concluding Remarks}
In the present letter we have described a development of the 
technique introduced in \cite{solov94a,solov94b} which
allows significantly greater control over the stucture of induced 
non-perturbative corrections. This has allowed an application
to $\ta$--decay and the resulting normalisation of the
effective coupling results in very good agreement with experimental
data at the $Z$--scale. 

The intriguing possibility raised by this technique is that, as discussed
in section 2, provided one assumes convergence of the variational series,
the induced power corrections will be consistent with the existence
of vacuum condensate operators in the OPE. Thus one may wonder
whether this formalism may allow an investigation of the meson spectrum
in the framework of QCD sum rules which does not explicitly 
include arbitrary condensate parameters (see e.g. \cite{dok95,dok96a}), 
the assumption being that these corrections,
at least in some averaged sense, are automatically induced by convergence
of the series. The results of an investigation of this kind will
be presented in a forthcoming publication \cite{jones97b}.

The authors thank D.V. Shirkov for discussions, and the 
support of A.R. by the Commonwealth Scholarship Commission and the British
Council, and the partial support of I.L.S. by a Royal Society Grant 
is gratefully acknowledged.

\bibliographystyle{prsty}

\end{document}